\documentclass[12pt]{article}

\usepackage{amsfonts,graphicx,caption,subcaption,amsmath,amssymb,
enumitem,url,authblk,amsthm,multicol,cite,float,bigints,tabularx}

\newtheorem{definition}{Definition}[section]
\newtheorem{lemma}{Lemma}[section]
\newtheorem{theorem}{Theorem}[section]
\newtheorem{remark}{Remark}[section]

\title{Mathematical model for acid water neutralization with anomalous and fast diffusion}
\author[1,2]{A. Ceretani}
\author[1,2]{J. Bollati}
\affil[1]{{\small Conicet-Universidad Austral,
Paraguay 1950, S2000FZF Rosario, Argentina}}
\affil[2]{{\small Facultad de Ciencias Exactas, Ingenier\'ia y Agrimensura,
Universidad Nacional de Rosario,
Pellegrini 250, S2000BTP Rosario, Argentina}}
\author[3]{L. Fusi}
\author[3]{F. Rosso}
\affil[3]{{\small Dipartimento di Matematica e Informatica ``U.Dini'',
Universit\`{a} degli Studi di Firenze,
Viale Morgagni 67/a, 50134 Firenze, Italy}}
\date{}

\begin{document}

\maketitle

\begin{abstract}
In this paper we model the neutralization of an acid solution in which 
the hydrogen ions are transported according to Cattaneo's diffusion. The latter is a modification of classical Fickian
diffusion in which the flux adjusts to the gradient with a positive 
relaxation time. Accordingly the evolution of the ions concentration 
is governed by the hyperbolic telegraph equation instead of the classical heat equation. We focus on the specific case of a marble slab reacting with a sulphuric acid solution and we consider a one-dimensional geometry. We show that the problem is multi-scale in time, with a reaction time scale that is larger than the diffusive time scale, so that the governing equation is reduced to the one-dimensional wave equation. The mathematical problem 
turns out to be a hyperbolic free boundary problem where the consumption of the slab is described by a nonlinear differential equation. Global well posedness is proved and some numerical simulations are provided.
\end{abstract}

%\begin{keyword} 

%% PACS codes here, in the form: \PACS code \sep code

%% MSC codes here, in the form: \MSC code \sep code
%% or \MSC[2008] code \sep code (2000 is the default)
%\end{keyword}
%\end{frontmatter}

%% \linenumbers

%% main text

%opening

\section{Introduction}\label{Sect:Intro}
When sulphide minerals present in rocks are exposed to air and water, sulphuric acid is produced as a consequence of a chemical reaction. The acidic water flow is known as Acid Rock Drainage (ARD). Though this is a natural process, it is highly enhanced by mining activities and it is then known as Acid Mine Drainage (AMD). AMD can cause serious damage in biodiversity and human health, especially after a mine plant has ended its activity. So treatment of acid water becomes a challenge in order to avoid long term environmental damage. A survey on remediation options for AMD can be found in \cite{JoHa2005}. The main techniques are based on chemical reactions neutralizing the acid water. A typical approach is the so-called limestone neutralization, which consists in the addition of a calcium carbonate base to the acidic water in order to reduce the acidity of the solution. In the last few years several models have been developed to describe the evolution of this sort of neutralizing systems \cite{FuMoPr2012,FuMoFaPr2013,FuFaPr2014, FuPrMo2015}. Although these models were proposed taking into account different aspects of the diffusive-reactive process, all of them are based on Fick's law (classical diffusion). When using Fick's law, it is tacitly assumed that local disturbances are spread infinitely fast throughout the solution. This is a clear idealization and it is physically unrealistic, giving rise to the pathological feature of infinitely fast spreading of perturbations in the diffusion equation. For this reason, Cattaneo and others have proposed to modify Fick's law in a way such that the flux may adjust to the gradient with a small but nonzero relaxation time, see \cite{Ca1948}. 

In this article we study the evolution of a neutralization process for an acid solution in which transport is driven by Cattaneo's law (anomalous diffusion). Following \cite{FuPrMo2015}, we consider the reaction occurring between a sulphuric acid solution ($\mathrm{H}_2\mathrm{SO}_4$) and a slab of marble, which is mainly formed by calcium carbonate ($\mathrm{CaCO}_3$):

\begin{equation}\label{Form:Chem}
\textrm{CaCO}_3\;\;+\;\; 2\textrm{H}^+\;\;+\;\;\textrm{SO}_4^{2-}\;\;\;\;\rightleftharpoons\;\;\;\;\textrm{Ca}^{2+}\;\;+\;\;\textrm{SO}_4^{2-}\;\;\;\;+\;\;\textrm{H}_2\textrm{O}\;\;+\;\;\textrm{CO}_2.
\end{equation} 

The $\mathrm{H}_2\mathrm{SO}_4$ dissolved in the aqueous solution is dissociated in ions $\textrm{SO}_4^{2-}$ and 2H$^+$, where the concentration of the latter is commonly measured through the so-called $\mathrm{pH}$. The acid reacts with $\mathrm{CaCO}_3$ liberating $\mathrm{Ca}^{2+}$ ions in the solution. As the reaction takes place, the $\mathrm{CaCO}_3$ is consumed and the concentration of $\mathrm{H}^{+}$ ions decreases (the $\mathrm{pH}$ is hence raised). When the  $\mathrm{pH}$ reaches a value around 7 the solution is said to be neutral and the reaction ceases. The reaction between $\mathrm{H}_2\mathrm{SO}_4$ and $\mathrm{CaCO}_3$ takes place on the contact surface that separates the solid and the solution. From (\ref{Form:Chem}) we see that the stoichiometric ratio in the reaction between $\mathrm{H}^{+}$ ions and $\mathrm{CaCO}_3$ is 2:1, which means that two moles of $\mathrm{H}^{+}$ are neutralized by one mole of $\mathrm{CaCO}_3$. Moreover the reaction is of first order, so that the exponents of the  concentrations appearing in the rate equation are equal to 1. When other reactants and/or acid solutions are used, different ratios and different reaction orders may occur. Here, differently from \cite{FuPrMo2015}, we do not consider the phenomenon of ``armoring'', that consists in the formation of a thin coat of material on the reacting surface which partially or completely inhibits the reaction.

In this paper we model a system consisting in a rectangular container filled with sulphuric acid where a slab of marble has been placed on the bottom. This slab is assumed to occupy less than the half of the container height at the beginning. We also assume that the container is large enough, so that the system can be described in a one dimensional geometry \footnote{Throughout the paper the starred quantities denote dimensional quantities.} $[0,L^*]$ where the solid occupies a region $[0,s^*]$, the liquid fills the region $[s^*,L^*]$, with $s^*$ evolving with time.  

We study the system for the evolution of both the  $\mathrm{H}^{+}$ concentration and the reacting surface $s^*$. We assume that no acid is added or removed during the process and we notice that our system is physically consistent only if and $\dot{s}^*<0$ (the solid slab can only be consumed). The transport of $\mathrm{H}^{+}$ in the solution is governed by Cattaneo's diffusion and the nature of the mathematical problem is therefore hyperbolic. The system is multi-scale in time with three characteristic times: i) the characteristic diffusive time; ii) the relaxation time; iii) the reaction time. Depending on the order of magnitude of these times different problems may arise. In particular, following \cite{DeSpPe2000} we show that the reaction scale is larger than the diffusion time scale, so that the consumption of the marble slab is slower than the diffusive transport of H$^{+}$ ions. Assuming that the relaxation time is sufficiently large we prove that the mathematical problem can be simplified to one in which the governing equation is the one dimensional wave equation. Following \cite{FuFa2003} we determine representation formulas for the solution that allow to write the evolution of the free boundary $s^*$ as an implicit nonlinear differential equation. Global existence and uniqueness are proved. Finally numerical simulations that illustrate the behavior of the solution $s^*$ and the dependence on the physical parameters of the problem are provided.

\section{Derivation of the model}\label{Sect:Derivation}
In this Section we derive the general model for the neutralization process described above. We consider a one-dimensional domain $[0,L^*]$ where $[0,s^*]$ is the region occupied by the reacting solid ($\mathrm{CaCO}_3$), $[s^*,L^*]$ is the region occupied by the acid solution ($2\mathrm{H}^{+}$ + $\textrm{SO}_4^{2-}$) and $s^*=s^*(t^*)$ is the free surface separating them. The $\mathrm{H}^{+}$ ions concentration in the solution will be denoted with $c^*=c^*(x^*,t^*)$ ([$c^*$]=$\mathrm{mol/length^3}$). 

\begin{remark}
The quantity $c^*$ provides a measure of the acidity of the solution. 
Alternatively one can use the so-called $\mathrm{pH}$ expressed as 
\begin{equation*}
\mathrm{pH}\ =-\log_{10}\left(\frac{c^*}{1\ \mathrm{mol}/\mathrm{lt}}\right).
\end{equation*}
The solution is said to be neutral when $\mathrm{pH}=7$ so that, in terms of $c^*$, we have that the solution is neutral when 
$c^*=10^{-7}\ \mathrm{mol}/\mathrm{lt}$.
\end{remark}
 
During the neutralization process the $\mathrm{H}^{+}$ ions diffuse in the liquid region $[s^*,L^*]$. We assume that diffusion is governed by the Cattaneo's law  
\begin{equation}\label{Cattaneo}
J^*+\tau^*\frac{\partial J^*}{\partial t^*}=-D^*\frac{\partial c^*}{\partial x^*},
\end{equation}  
where $J^{*}$ is the ions flux, $\tau^*>0$ is a relaxation time and $D^*$ is the diffusivity coefficient ([$D^*$]=$\mathrm{length}^2/\mathrm{time}$), which is assumed to be constant.

\begin{remark}
When $\tau^*\rightarrow 0^+$, the process becomes purely diffusive and (\ref{Cattaneo}) reduces to Fick's law
\begin{equation*}
J^*=-D^*\frac{\partial c^*}{\partial x^*}.
\end{equation*}
In Cattaneo's law the flux is allowed to adjust to the gradient of concentration according to a relaxation time $\tau^*$. In fact, it can be seen as an approximation of the constitutive equation
\begin{equation}\label{Form:J}
J^*(x^*,t^*+\tau^*)=-D^*\frac{\partial c^*}{\partial x^*}(x^*,t^*).
\end{equation}
Moreover the flux is mainly influenced by what has happened close in time to present. In effect, the flux $J^*$ can be explicitly written as
\begin{equation}\label{ShortTail}
J^*(x^*,t^*)=-D^*\displaystyle\int_{-\infty}^{t^*}K^*(t^*-\widetilde{t^*})\frac{\partial c^*}{\partial x^*}(x^*,\widetilde{t^*})d\widetilde{t^*},
\end{equation}
where $K^*$ is the short-tail kernel given by
\begin{equation*}
K^*(t^*)=\frac{1}{\tau^*}\exp\left(-\frac{t^*}{\tau^*}\right),
\end{equation*}
which rapidly decays to zero. When other types of kernels are considered, the system may express a different memory behavior, see for example \cite{GoMa1997}, where long-tail kernels are used in the context of fractional diffusion, or \cite{GuPi1968} where more general kernels are considered.     
\end{remark}

It is easy to show that the continuity equation
\begin{equation}\label{Eq:c_Cont}
\frac{\partial c^*}{\partial t^*}=-\frac{\partial J^*}{\partial x^*},
\end{equation}
together with Cattaneo's law (\ref{Cattaneo}), provides the following telegraph equation for $c^*$:
\begin{equation}\label{Eq:c_Telegraph}
\frac{\partial c^*}{\partial t^*}+\tau^*\frac{\partial^2 c^*}{\partial {t^*}^2}=D^*\frac{\partial^2 c^*}{\partial {x^*}^2}.
\end{equation}
Equation (\ref{Eq:c_Telegraph}) is the governing equation for the evolution of the ions concentration in the solution. 
Differently from the case of pure Fickian diffusion in which the governing equation is parabolic, here the nature of the problem is hyperbolic.  

To describe the consumption of the solid part $[0,s^*]$ we consider the rate equation that governs the chemical reaction occurring on the free boundary $s^*$. Following \cite{FuFaPr2014,FuPrMo2015} we write 
\begin{equation}\label{me}
v^*=-k^*(c^*-c^*_0)_+,
\end{equation}  
where $v^*$ represents the rate of neutralized $\mathrm{H}^{+}$ moles per unit surface, $k^*$ is the reaction rate and $c^*_0$ is the concentration of neutralization \footnote{The dimensions of the quantities in (\ref{me}) are: [$v^*$]=$\mathrm{mol}/(\mathrm{length}^2\cdot \mathrm{time})$;  [$k^*$]=$\mathrm{length}/\mathrm{time}$; [$c^*_0$]=$\mathrm{mol}/\mathrm{length}^3$.}. The velocity of the reaction is therefore proportional to the excess of ions on the reacting surface. The positive part is taken to prevent the reaction from occurring when the concentration is above the neutralization limit.

Assuming that the ``molar'' density \footnote{To avoid dimensional inconsistencies we assume that the marble density $\rho^*$ is the molar density, i.e. classical density divided by the molecular weigth.} of  $\mathrm{CaCO}_3$ is constant we write
\begin{equation*}
\frac{d}{dt^*}\left(\displaystyle\int_0^{s^*}\rho^*dx^*\right)=v^*,
\end{equation*}   
from which follows that
\begin{equation}\label{Cond:s_FreeBound-1}
\rho^*\dot{s}^*=-k^*(c^*-c^*_0)_+.
\end{equation}
This condition on the free boundary $s^*$ implies that $\dot{s}^{*}$ must be negative while $c^{*}>c_0^{*}$. This is in agreement with the fact that $\mathrm{CaCO}_3$ is consumed when the $\mathrm{H}^{+}$ concentration is greater than $c_0^*$. Recalling that the stoichiometric ratio is 2:1, the overall mass balance is given by 
\begin{equation*}
2\frac{d}{dt^*}\left(\displaystyle\int_{s^*}^{L^*}c^*dx^*\right)=
\frac{d}{dt^*}\left(\displaystyle\int_0^{s^*}\rho^*dx^*\right),
\end{equation*}
yielding 
\begin{equation*}
2\displaystyle\int_{s^*}^{L^*}\frac{\partial c^*}{\partial t^*}dx^*-2c^*(s^*,t^*)\dot{s}^*=\rho^*\dot{s}^*.
\end{equation*}
Exploiting (\ref{Eq:c_Cont}) we find 
\begin{equation*}
2\Big[J^*(s^*,t^*)
-J^*(L^*,t^*)\Big]=
\dot{s}^*\left(\rho^*+2c^*(s^*,t^*)\right).
\end{equation*}
Since $\mathrm{H}^{+}$ ions are not added or removed at $x^*=L^*$, it is reasonable to impose the boundary condition $J^*(L^*,t^*)=0$ so that
\begin{equation}\label{Form:J}
2J^*(s^*,t^*)=\dot{s}^*\left(\rho^*+2c^*(s^*,t^*)\right).
\end{equation} 
Differentiating the last expression with respect to time we find
\begin{equation*}
2\dot{s}^*\frac{\partial J^*}{\partial x^*}(s^*,t^*)+
2\frac{\partial J^*}{\partial t^*}(s^*,t^*)=2\dot{s^*}\left(\dot{s}^*\frac{\partial c^*}{\partial x^*}(s^*,t^*)+\frac{\partial c^*}{\partial t^*}(s^*,t^*)\right)+
\end{equation*}
\begin{equation*}
+\ddot{s}^*\left(\rho^*+2c^*(s^*,t^*)\right).
\end{equation*}
Hence, recalling (\ref{Cattaneo}), (\ref{Eq:c_Cont}), we get
\begin{equation*}
\begin{split}
-2\dot{s}^*\frac{\partial c^*}{\partial t^*}(s^*,t^*)&-
\frac{2}{\tau^*}\left(D^*\frac{\partial c^*}{\partial x^*}(s^*,t^*)+J^*(s^*,t^*)\right)=\\[0.2cm]
&2\dot{s^*}\left(\dot{s}^*\frac{\partial c^*}{\partial x^*}(s^*,t^*)+\frac{\partial c^*}{\partial t^*}(s^*,t^*)\right)+
\ddot{s}^*\left(\rho^*+2c^*(s^*,t^*)\right).
\end{split}
\end{equation*}
Finally, taking into account (\ref{Form:J}) we find
\begin{equation*}
-2D^*\frac{\partial c^*}{\partial x^*}(s^*,t^*)=
\dot{s}^*\left(\rho^*+2c^*(s^*,t^*)\right)+
\end{equation*}
\begin{equation}\label{Cond:s_FreeBound-2}
+\tau^*\left[\ddot{s}^*\left(\rho^*+2c^*(s^*,t^*)\right)
+2{\dot{s}^*}{^2}\frac{\partial c^*}{\partial x^*}(s^*,t^*)
+4\dot{s}^*\frac{\partial c^*}{\partial t^*}(s^*,t^*)\right].
\end{equation}

The mathematical formulation of our model is therefore a hyperbolic free boundary problem consisting 
of the  telegraph equation (\ref{Eq:c_Telegraph}) on $[s^*,L^*]$, and the conditions (\ref{Cond:s_FreeBound-1}), (\ref{Cond:s_FreeBound-2}) to which we must add the initial data
%\footnote{The condition on $\dot{s}_0^*$ is actually redundant, since it can be derived from (\ref{Cond:s_FreeBound-1}).}
\begin{align}
\label{Cond:c_Initial}&c^*(x^*,0)=c^*_{in_0}(x^*),&\quad 
&\frac{\partial c^*}{\partial t^*}(x^*,0)=c^*_{in_1}(x^*)&\quad
&s^*_0\leq x^*\leq L^*,\\
\label{Cond:s_Initial}&s^*(0)=s^*_0,&\quad 
&\dot{s}^*(0)=\dot{s}_0^*,&\quad
&
\end{align}
where $0<s^*_0\leq L^*/2$, and the zero-flux boundary condition:
\begin{equation}\label{Cond:c_Boundary}
\frac{\partial c^*}{\partial x^*}(L^*,t^*)=0\hspace*{1cm}0<t^*.
\end{equation}
Such a problem will be referred to as (P$^*_c$). 

\section{The non-dimensional formulation}\label{Sect:Non-Dimensions}
To investigate the multi-scale nature of problem (P$^*_c$) it is convenient to rewrite it in a non-dimensional form. For this purpose, we introduce the characteristic times
\begin{equation*}
t^*_D=\frac{{L^*}^2}{D^*},\hspace*{2cm}
t^*_R=\frac{L^*}{k^*},
\end{equation*}
representing the diffusion characteristic time and the reaction characteristic time, \mbox{respectively}. We also introduce the reference concentration $c^*_A=\max_{[s^*_0,L^*]} c^*_{in_0}(x^*)$  and the non-dimensional parameters
\begin{equation*}
\lambda=\frac{\rho^*}{2 c^*_A},\hspace*{2cm}
\delta=\frac{c^*_0}{c^*_A}.
\end{equation*}
Then we rescale the main variables as follows 
\begin{equation*}
c=\frac{c^*}{c^*_A},\quad
t=\frac{t^*}{t^*_{ref}},\quad
x=\frac{x^*}{L^*},\quad
s=\frac{s^*}{L^*},
\end{equation*}
where $t_{ref}^*$ is a reference time to be selected. The non-dimensional version of problem (P$^*_c$) becomes

\begin{equation}\label{Prob:c_Telegraph}
(\text{P}_c)\,:\,\left\{
\begin{array}{lll}
\left(\dfrac{t^*_D}{t^*_{ref}}\right)\,c_t+
\left(\dfrac{t^*_D\tau^*}{{t^*_{ref}}^2}\right)\,c_{tt}=c_{xx}
&s<x<1,\,0<t\\
\\
c(x,0)=c_{in_0}(x)&s_0<x<1\\
\\
c_t(x,0)=c_{in_1}(x)&s_0<x<1\\
\\
c_x(1,t)=0&0<t\\
\\
2\lambda\dot{s}=-\left(\dfrac{t^*_{ref}}{t^*_R}\right)\left(c(s,t)-\delta\right)_+
&0<t\\
\\
-c_x(s,t)=\left(\dfrac{t^*_D}{t^*_{ref}}\right)\,\dot{s}\left(c(s,t)+\lambda\right)+&&\\
\\
+\,\left(\dfrac{t^*_D\tau^*}{{t^*_{ref}}^2}\right)\left(\ddot{s}\left(c(s,t)+\lambda\right)+\dot{s}^2c_x(s,t)+
2\dot{s}c_t(s,t)\right)&0<t\\
\\
s(0)=s_0&&\\
\\
\dot{s}(0)=\dot{s}_0,&&
\end{array}\right.
\end{equation}
where
\begin{equation*}
c_{in_0}=\frac{c^*_{in_0}}{c^*_A},\quad
c_{in_1}=\frac{t^*_{ref}}{c^*_A}c^*_{in_1},\quad
s_0=\frac{s^*_0}{L^*},\quad
\dot{s}_0=\frac{t^*_{ref}}{L^*}\dot{s}^*_0,
\end{equation*}
and $0<s_0\leq 1/2$.

We notice that there are three characteristic times in problem (P$_c$). One given by the characteristic time of diffusion $t^*_D$, another given by the characteristic time of the chemical reaction $t^*_R$, and one more given by the relaxation time $\tau^*$.

\begin{remark}
When $\tau^*\to 0^+$ equation (\ref{Prob:c_Telegraph})$_1$ becomes the diffusion equation and condition (\ref{Prob:c_Telegraph})$_6$ reduce to that formulated in \cite{FuPrMo2015}, where the pure diffusive case was studied.
\end{remark}

When considering $\mathrm{CaCO}_3$ in an acid  solution $\mathrm{H}_2\mathrm{SO}_4$, typical values are:
\begin{equation*}
\rho^*=2.7 \cdot 10^{-2}\text{ mol/lt},\quad
c^*_0\sim 10^{-7}\text{ mol/lt},\quad
c^*_A=10^{-2}\text{ mol/lt}.
\end{equation*}
Hence 
\begin{equation*}
\lambda\simeq 1.3,\quad \delta\sim 10^{-5}.
\end{equation*}

\section{Fast diffusion and slow relaxation}\label{Sect:ReactionScale}
Following \cite{DeSpPe2000,LeRa2011} we consider the typical values
\begin{equation*}
D^*=5\cdot 10^{-5}\ \ cm^2/s, \ \ \ \ \ \ \ \ L^*=10\ \ cm, \ \ \ \ \ \ \ \ \ \ \ \ k^*=10^{-7} \ \ cm/s,
\end{equation*}
so that 
\begin{equation*}
\dfrac{t^*_D}{t^*_R}=0.02=O(10^{-2}).
\end{equation*}
The above means that the diffusion is faster than reaction. Therefore, assuming that relaxation is very slow
and choosing the reaction time as reference time $t^*_{ref}=t^*_R$, we can write 
\begin{equation}\label{deh}
\dfrac{t^*_D}{t^*_R}\ll 1, \ \ \ \ \ \ \ \ \ \ \ \ \ \ \dfrac{t^*_D\tau^*}{{t^*_R}^2}=\mathcal{O}(1),
\end{equation}
which clearly implies 
\begin{equation*}
t^*_D\ll t^*_R \ll \tau^*.
\end{equation*}
In conclusion we have that the diffusive scale is faster than the reaction scale which, in turn, is faster than the relaxation scale. When (\ref{deh}) holds, the terms containing  $(t^*_D/ t^*_R)$ in (\ref{Prob:c_Telegraph})$_1$, (\ref{Prob:c_Telegraph})$_6$ can be safely neglected and the problem (P$_c$) reduces to
\begin{equation}\label{Prob:c_Wave}
(\widetilde{\text{P}}_c)\,:\,
\left\{\begin{array}{ll}
c_{tt}=\alpha^2\,c_{xx}&s<x<1,\,0<t\\
\\
c(x,0)=c_{in_0}(x)&s_0<x<1\\
\\
c_t(x,0)=c_{in_1}(x)&s_0<x<1\\
\\
c_x(1,t)=0&0<t\\
\\
2\lambda\dot{s}=-\left(c(s,t)-\delta\right)_+&0<t\\
\\
-\alpha^2\,c_x(s,t)=\ddot{s}\left(c(s,t)+\lambda\right)+\dot{s}^2c_x(s,t)+
2\dot{s}c_t(s,t)&0<t\\
\\
s(0)=s_0\\
\\
\dot{s}(0)=\dot{s}_0,
\end{array}\right.
\end{equation}
where we have set 
$$\alpha^2=\frac{{t^*_R}^2}{t^*_D\tau^*}=\mathcal{O}(1).$$

In this peculiar situation the governing equation is the one-dimensional wave equation, whose solution can be expressed by means of D'Alembert formulas.

The remainder of this Section is devoted to proving the existence and uniqueness of a solution $(c,s)$ to problem ($\widetilde{\text{P}}_c$) according to the following definition:

\begin{definition}\label{Def:c_WaveSolution}
The pair $(c,s)$ is said to be a classical solution to problem ($\widetilde{\text{P}}_c$) in the time interval $[0,T]$ if:
\begin{enumerate}
\item[i)] $c$, $s$ are defined on $\overline{D_T}$ and $[0,T]$ respectively;
\item[ii)] $c\in C^{2,2}(\overline{D}_T)$
%\footnote{Actually the second derivatives of $c(x,t)$ can be discontinuous on the characteristic line $x+\alpha t=s_0$, depending on the initial data.} 
and $s\in C^{2}[0,T]$; 
\item[iii)] $c$, $s$ satisfy ($\widetilde{\text{P}}_c$);
\item[iv)] $-\alpha<\dot{s}<0$ (the solid can only be consumed by the reaction);
\end{enumerate} 
where  $D_T$ is the liquid domain:
\begin{equation*}
D_T=\left\{(x,t)\in\mathbb{R}^2\,:\,s(t)<x<1,\,0<t<T\right\}.
\end{equation*}
\end{definition}
\begin{remark}
From the Definition \ref{Def:c_WaveSolution} it follows that
equation (\ref{Prob:c_Wave})$_{5}$ must hold up to time $t=0$, so that
\begin{equation*}
\dot{s}_0=-\frac{1}{2\lambda}\left(c_{in_0}(s_0)-\delta\right)_+
\end{equation*}
is a necessary condition that must be fulfilled by any classical solution.
\end{remark}

\subsection{Representation formulas}\label{Sect:RepresentForm}
Suppose for a moment that problem (\ref{Prob:c_Wave}) has a unique solution $(c,s)$ in the sense of Definition \ref{Def:c_WaveSolution} on some time interval $[0,T]$. An explicit expression for $c$ in terms of $s$ and of the data of problem ($\widetilde{\text{P}}_c$) can be determined exploiting D'Alembert fundamental formula. Following \cite{FuFaPr2014} we see that it is convenient to transform the problem ($\widetilde{\text{P}}_c$) into a Stefan-like problem, in order to make the free boundary conditions more manageable. Considering the transformation
\begin{equation}\label{c-u_Transformation}
u(x,t)=\lambda(x-s(t))+\displaystyle\int_s^xc(\eta,t)d\eta,
\end{equation}
the problem for the new variable $u$ becomes
\begin{equation}\label{Prob:u_Wave}
(\widetilde{\text{P}}_u)\,:\,
\left\{
\begin{array}{ll}
u_{tt}=\alpha^2\,u_{xx}&s<x<1,\,0<t\\
\\
u(x,0)=u_{in_0}(x)&s_0<x<1\\
\\
u_t(x,0)=u_{in_1}(x)&s_0<x<1\\
\\
u(1,t)=\mu(t)&0<t\\
\\
u(s,t)=0&0<t\\
\\
2\lambda\dot{s}=-\left(u_x(s,t)-\lambda-\delta\right)_+&0<t\\
\\
s(0)=s_0,
\end{array}\right.
\end{equation}
where $u_{in_0}$, $u_{in_1}$, $\mu$ are the functions given by
\begin{equation}\label{u_Data}
\begin{split}
&u_{in_0}(x)=\lambda(x-s_0)+\displaystyle\int_{s_0}^xc_{in_0}(\eta)d\eta,\\
&u_{in_1}(x)=-\dot{s}_0(\lambda+c_{in_0}(s_0))+\displaystyle\int_{s_0}^xc_{in_1}(\eta)d\eta,\\
&\mu(t)=u_{in_1}(1)t+u_{in_0}(1),
\end{split}
\end{equation}
and $\dot{s}_0$ is defined by
\begin{equation}\label{Cond:s1_Compatibility-1}
\dot{s}_0=-\frac{1}{2\lambda}\left(u'_{in_0}(s_0)-\lambda-\delta\right)_+.
\end{equation} 
Once ($\widetilde{\text{P}}_u$) is solved we may get the solution of ($\widetilde{\text{P}}_c$)  through (\ref{c-u_Transformation}). The main advantage of ($\widetilde{\text{P}}_u$) lies in the free boundary conditions. 

\begin{remark} The function $\mu(t)$ can be obtained in the following way. Observe that 
$u_{xx}=c_x=\alpha^{-2}u_{tt}$ everywhere in $\overline{D}_T$. As a consequence $c_x(1,t)=\alpha^{-2}u_{tt}(1,t)=0$ and 
$u_t(1,t)=const$. Imposing the compatibility conditions on $x=1$ we find that $u_t(1,t)=u_t(1,0)=u_{in_1}(1)$ and, integrating in $t$, we finally find 
$$u(1,t)= u_{in_1}(1)t+u_{in_0}(1)=\mu(t).$$
\end{remark}

\begin{figure}
\centering
\includegraphics[scale=0.4]{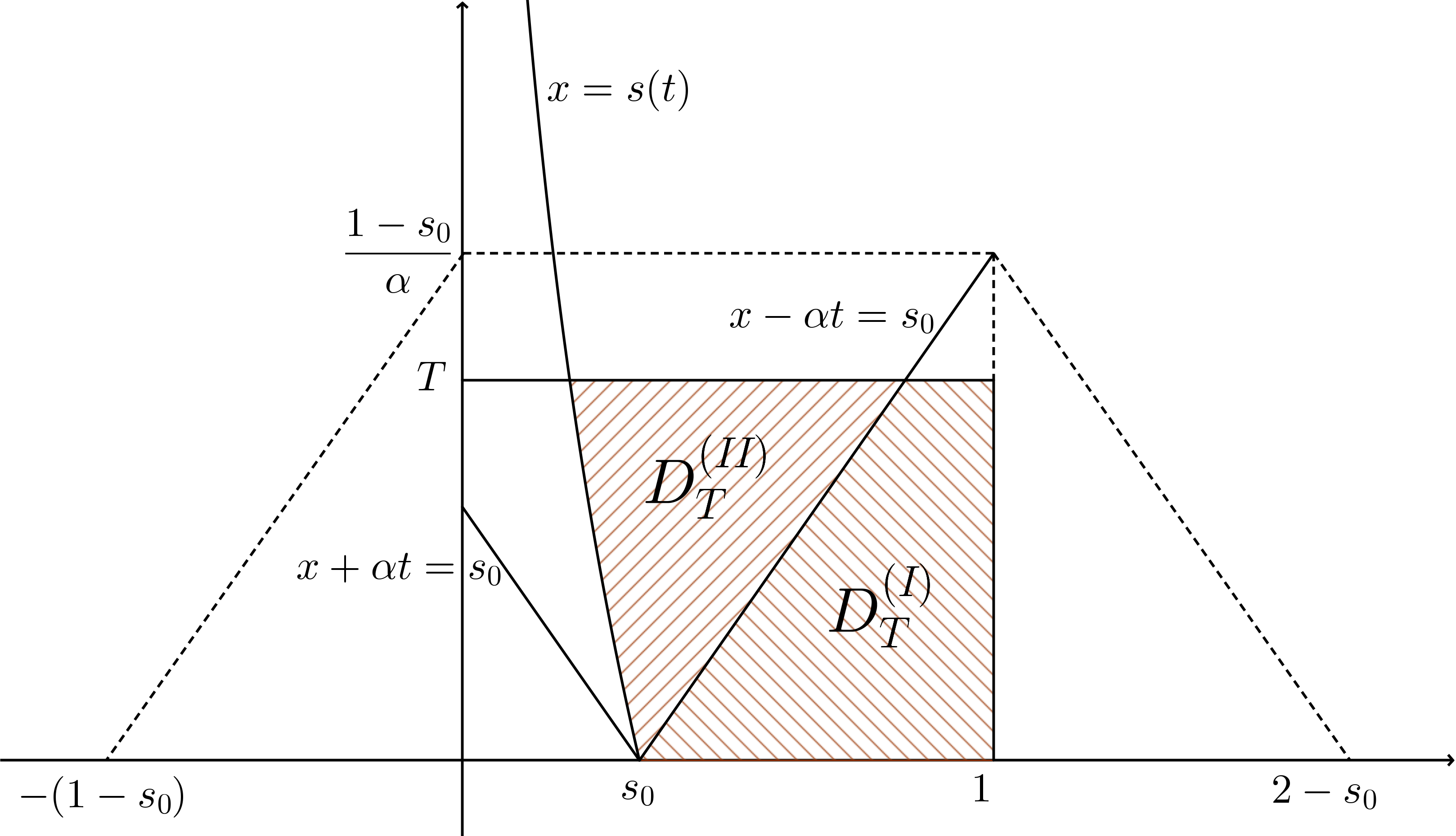}
\caption{Schematic representation of the liquid domain $D_T$ split into $D_T^{(I)}$ and $D_T^{(II)}$.}
\label{Fig:DomainD}
\end{figure}

Let $(u,s)$ be the classical solution of ($\widetilde{\text{P}}_u$) related to $(c,s)$ according to (\ref{c-u_Transformation}). In order to use D'Alembert representation formulas for $u$, we split the liquid domain $D_T$ as (Fig. \ref{Fig:DomainD})
\begin{align*}
&D_{T}^{(I)}=\left\lbrace (x,t)\in D_T: s_0+\alpha t<x<1, 0<t<T\right\rbrace, \\[0.2cm]
&D_{T}^{(II)}=\left\lbrace (x,t)\in D_T: s(t)<x<s_0+\alpha t, 0<t<T\right\rbrace. 
\end{align*}
Recalling that $s_0\leq1/2$, the characteristics curves emerging from $(s_0,0)$ meet the external boundaries $x=0$, $x=1$ in 
$$ T_l=\dfrac{s_0}{\alpha}, \ \ \ \ \ \ \ \ \ \ \ \ \ \ \ \ \ \ \ \ \ \ T_r=\dfrac{1-s_0}{\alpha}, $$
respectively, with $T_l\leq T_r$. Looking at Fig. \ref{Fig:DomainD} we notice that it is natural to seek a solution in the time interval $[0,T]$ with
\begin{equation}\label{T}
0<T\leq T_r=\frac{1-s_0}{\alpha},
\end{equation}
and to consider a suitable extension of data to the interval $[-(1-s_0),2-s_0]$. Thus, we will set
\begin{equation*}
U_j(x)=\left\{
\begin{array}{lcl}
u_{in_j}(x)&\text{if }&s_0\leq x\leq 1\\
\\
\widetilde{u}_j(x)&\text{if }&1<x\leq 2-s_0\\
\\
\widehat{u}_j(x)&\text{if }&-(1-s_0)\leq x<s_0
\end{array} \right.\quad j=0,1,
\end{equation*}
where $\widetilde{u}_j$, $\widehat{u}_j$ are functions that must be determined. Then D'Alembert formula states that 
\begin{equation}\label{Form:uDI-1}
u(x,t)=\frac{U_0(x+\alpha t)+U_0(x-\alpha t)}{2}+
\dfrac{1}{2\alpha}\displaystyle\int\limits_{x-\alpha t}^{x+\alpha t}U_1(\eta)d\eta.
\end{equation}

\begin{figure}
\centering
\includegraphics[scale=0.4]{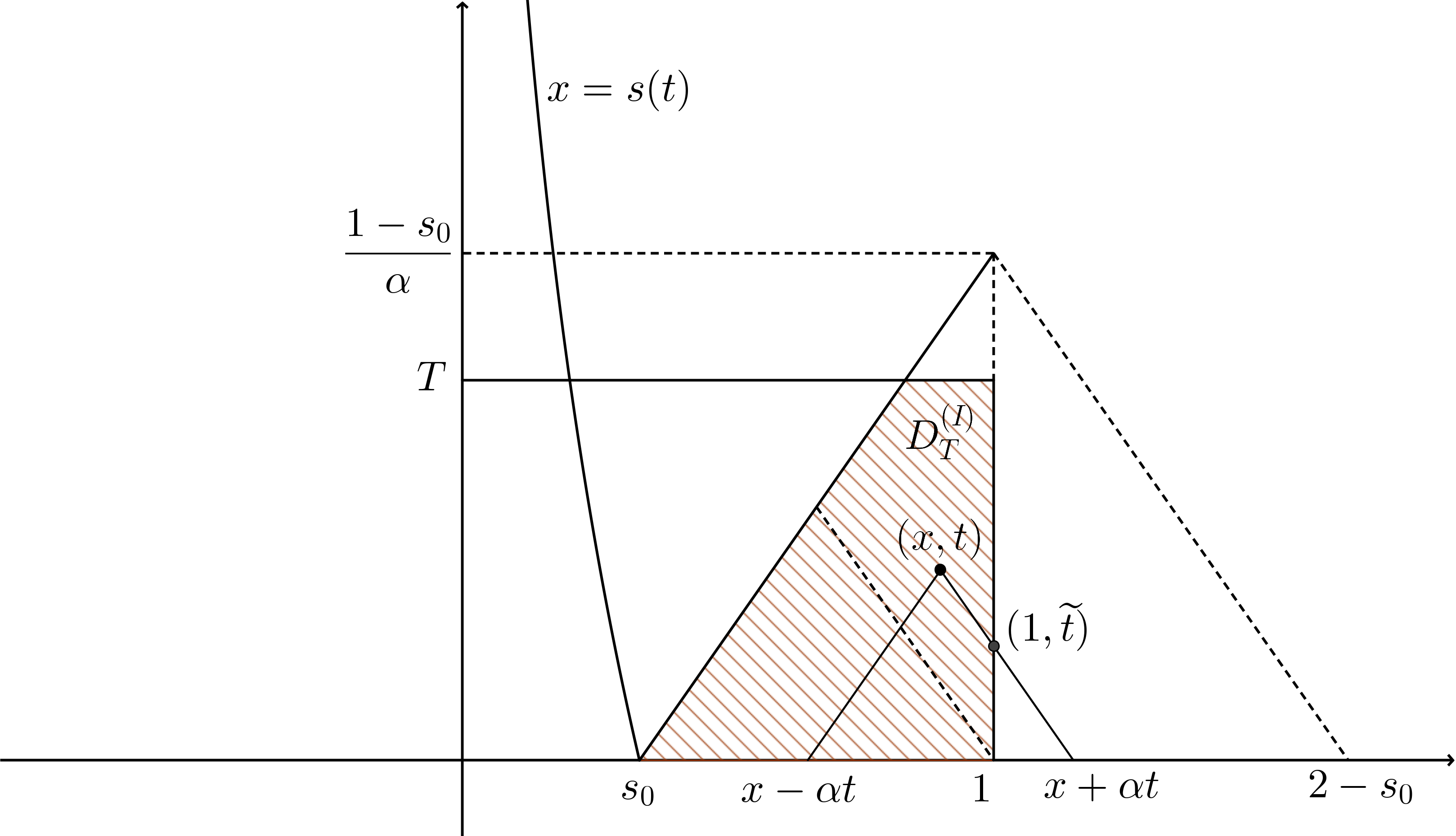}
\caption{Characteristics lines for $(x,t)\in D_T^{(I)}$.}
\label{Fig:Extensions-DI}
\end{figure}

$$$$
\noindent {\em Representation formula in $D_T^{(I)}$}
$$$$
\noindent In this first case, data $u_{in_0}$, $u_{in_1}$ only need to be extended to the right of $x=1$ (Fig. \ref{Fig:Extensions-DI}). Then only the functions $\widetilde{u}_{j}$ must be determined. We will define them in a way such that the boundary condition (\ref{Prob:u_Wave})$_4$ holds.

On one hand, if $x+\alpha t\leq 1$ formula (\ref{Form:uDI-1}) reduces to
\begin{equation}\label{Form:uDI-2}
u(x,t)=\dfrac{u_{in_0}(x+\alpha t)+u_{in_0}(x-\alpha t)}{2}+
\dfrac{1}{2\alpha}\displaystyle\int\limits_{x-\alpha t}^{x+\alpha t}u_{in_1}(\eta)d\eta.
\end{equation}

On the other hand, when $x+\alpha t>1$, (\ref{Form:uDI-1}) can be written as
\begin{equation}\label{Form:uDI-3}
u(x,t)=\dfrac{\widetilde{u}_0(x+\alpha t)+u_{in_0}(x-\alpha t)}{2}+
\dfrac{1}{2\alpha}\displaystyle\int\limits_{x-\alpha t}^{x+\alpha t}U_1(\eta)d\eta.
\end{equation}
Let $(1,\widetilde{t})$ be the point in which the characteristic curve joining the points $(x,t)$, $(x+\alpha t,0)$ meets the line $x=1$ (Fig. \ref{Fig:Extensions-DI}). From condition (\ref{Prob:u_Wave})$_4$ and formula (\ref{Form:uDI-3}), we obtain
\begin{equation}\label{Form:U0DI-1}
\mu(\widetilde{t})=\dfrac{\widetilde{u}_0(1+\alpha\widetilde{t})+
u_{in_0}(1-\alpha\widetilde{t})}{2}+
\dfrac{1}{2\alpha}\displaystyle\int\limits_{1-\alpha\widetilde{t}}^
{1+\alpha\widetilde{t}}U_1(\eta)d\eta.
\end{equation}
Noting that $x+\alpha t=1+\alpha \widetilde{t}$, it follows that
\begin{equation*}
\widetilde{t}=\dfrac{x+\alpha t-1}{\alpha}.
\end{equation*}
Combining this with (\ref{Form:U0DI-1}) we find
\begin{equation}\label{Form:U0DI-2}
\dfrac{\widetilde{u}_0(x+\alpha t)}{2}=
\mu\left(\dfrac{x+\alpha t-1}{\alpha}\right)-
\dfrac{u_{in_0}(2-x-\alpha t)}{2}-\frac{1}{2\alpha}\displaystyle\int\limits_{2-x-\alpha t}^{x+\alpha t}U_1(\eta)d\eta.
\end{equation}
Replacing (\ref{Form:U0DI-2}) in (\ref{Form:uDI-1}), we obtain 
\begin{equation*}
u(x,t)=\mu\left(\dfrac{x+\alpha t-1}{\alpha}\right)+\dfrac{u_{in_0}(x-\alpha t)-u_{in_0}(2-x-\alpha t)}{2}+
\dfrac{1}{2\alpha}\displaystyle\int\limits_{x-\alpha t}^{2-x-\alpha t}u_{in_1}(\eta)d\eta.
\end{equation*}
Therefore, the representation formula for $u$ in $D_T^{(I)}$ is
\begin{equation}\label{Form:uDI}
u(x,t)=w(x,t)+\dfrac{u_0(x-\alpha t)+u_0(x+\alpha t)}{2}+
\dfrac{1}{2\alpha}\displaystyle\int\limits_{x-\alpha t}^{x+\alpha t}u_1(\eta)d\eta,
\end{equation}
where we have set
\begin{equation}\label{w}
w(x,t)=\left\{\begin{array}{lcl}
\mu\left(\dfrac{x+\alpha t-1}{\alpha}\right)&\text{if}&x+\alpha t\geq 1\\
0&\text{if}&x+\alpha t<1,
\end{array}\right.
\end{equation}
and 
\begin{equation}\label{uj}
u_j(x)=\left\{\begin{array}{lcl}
u_{in_j}(x)&\text{if}&s_0\leq x\leq 1\\
\\
-u_{in_j}(2-x)&\text{if}&1< x\leq 2-s_0
\end{array}\right.\quad j=0,1.
\end{equation}

\begin{figure}
\centering
\includegraphics[scale=0.4]{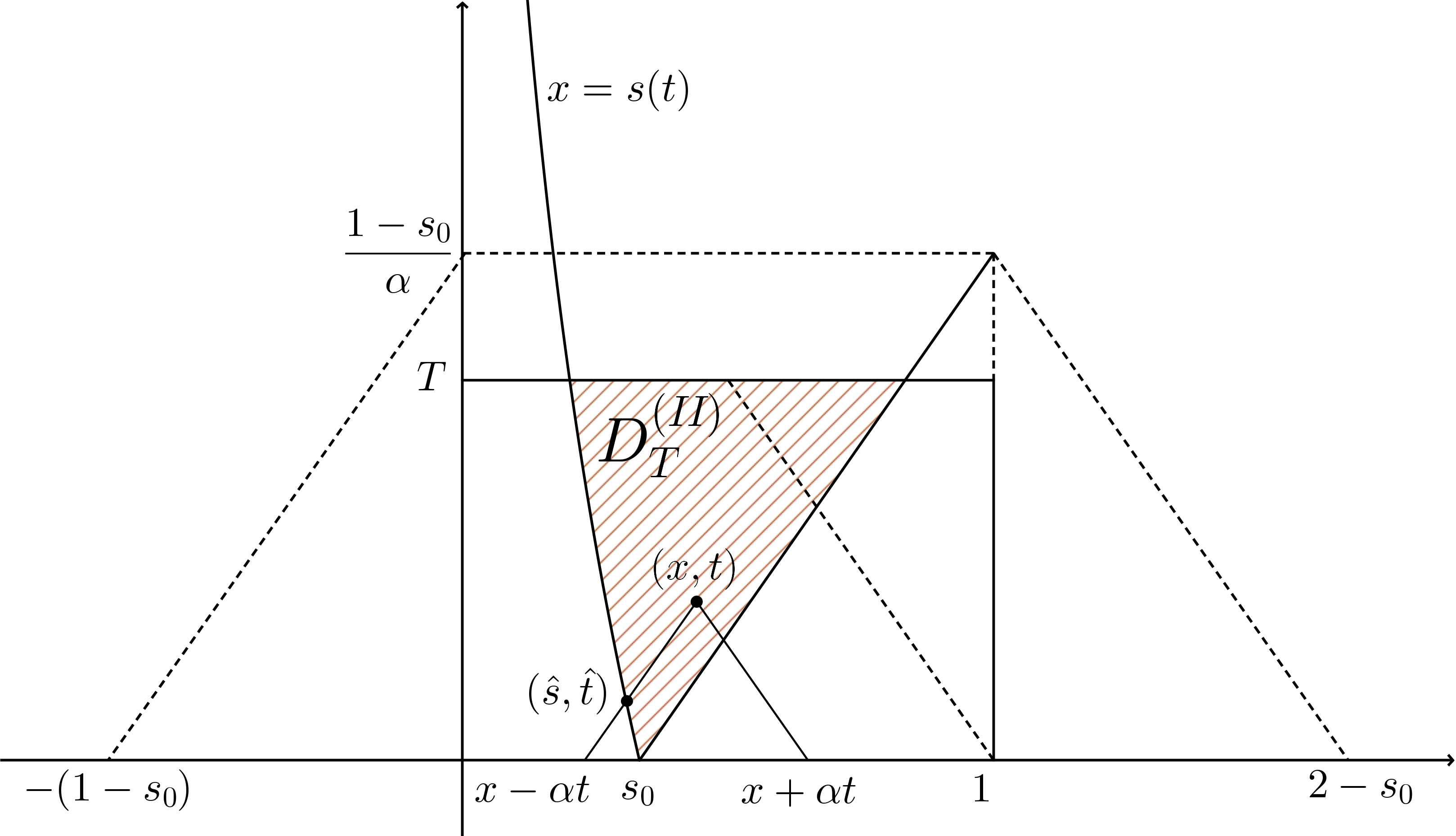}
\caption{Characteristics lines for $(x,t)\in D_T^{(II)}$.}
\label{Fig:Extensions-DII}
\end{figure}

\begin{remark}\label{remcont}
Notice that, even though the functions $w$, $u_0$, $u_1$ may be discontinuous at $x=1$ (this is true unless $u_{in_j}(1)=0$), the function $u(x,t)$ is continuous across the characteristic $x+\alpha t=1$. To prove this, it is sufficient to check from (\ref{Form:uDI}) that 
\begin{equation*}
\lim_{x+\alpha t\rightarrow 1^{+}}u(x,t)=\lim_{x+\alpha t\rightarrow 1^{-}}u(x,t).
\end{equation*}
\end{remark}
$$$$
\noindent {\em Representation formula in $D_T^{(II)}$}
$$$$
In this second case, data $u_{in_0}$, $u_{in_1}$ need to be extended to the right of $x=1$ as well as to the left of $x=s_0$ (Fig. \ref{Fig:Extensions-DII}). Therefore, functions $\widehat{u}_j$ will be also involved in D'Alembert formula. Similarly as before, we will define them in a consistent way with the condition (\ref{Prob:u_Wave})$_5$. 

When $x+\alpha t\leq1$, formula (\ref{Form:uDI-1}) becomes
\begin{equation}\label{Form:uDII-1}
u(x,t)=\dfrac{u_{in_0}(x+\alpha t)+\widehat{u}_0(x-\alpha t)}{2}+
\dfrac{1}{2\alpha}\displaystyle\int\limits_{x-\alpha t}^{x+\alpha t}U_1(\eta)d\eta.
\end{equation}
Let $\left(s(\hat{t}),\hat{t}\right)$ be the point in which the characteristic line joining $(x,t)$, $(x-\alpha t,0)$ meets the free boundary $s$ (Fig. \ref{Fig:Extensions-DII}). This point is uniquely defined because $\vert \dot{s}\vert<\alpha$. Moreover $\hat{t}=\hat{t}(x,t)$ is the unique solution to the equation
\begin{equation}\label{hat-t}
\hat{s}-\alpha \hat{t}=x-\alpha t,
\end{equation}
where we have set $\hat{s}=s(\hat{t})$.

From condition (\ref{Prob:u_Wave})$_5$ and formula (\ref{Form:uDII-1}), we find
\begin{equation*}
\dfrac{u_{in_0}(\hat{s}+\alpha \hat{t})+\widehat{u}_0(\hat{s}-\alpha \hat{t})}{2}+
\dfrac{1}{2\alpha}\displaystyle\int\limits_{\hat{s}-\alpha \hat{t}}^{\hat{s}+\alpha \hat{t}}U_1(\eta)d\eta=0.
\end{equation*}
Combining this with (\ref{hat-t}), we obtain
\begin{equation}\label{U0DII}
\dfrac{\widehat{u}_0(x-\alpha t)}{2}=-\dfrac{u_{in_0}(\hat{s}+\alpha\hat{t})}{2}-\frac{1}{2\alpha}\displaystyle\int\limits_{x-\alpha t}^{\hat{s}+\alpha \hat{t}}U_1(\eta)d\eta.
\end{equation}
Replacing (\ref{U0DII}) in (\ref{Form:uDII-1}), we find
\begin{equation}\label{Form:uDII-2}
u(x,t)=\dfrac{u_{in_0}(x+\alpha t)-u_{in_0}(\hat{s}+\alpha\hat{t})}{2}+\dfrac{1}{2\alpha}\displaystyle\int\limits_{\hat{s}+\alpha\hat{t}}^{x+\alpha t}u_{in_1}(\eta)d\eta.
\end{equation}
Finally, when $x+\alpha t>1$ formula (\ref{Form:uDI-1}) becomes
\begin{equation}\label{Form:uDII-3}
u(x,t)=\dfrac{\widehat{u}_0(x-\alpha t)+\widetilde{u}_0(x+\alpha t)}{2}+
\frac{1}{2\alpha}\displaystyle\int\limits_{x-\alpha t}^{x+\alpha t}U_1(\eta)d\eta.
\end{equation}
Replacing (\ref{Form:U0DI-2}) and (\ref{U0DII}) in (\ref{Form:uDII-3}) we find
\begin{equation}\label{Form:uDII-4}
u(x,t)=\mu\left(\dfrac{x+\alpha t-1}{\alpha}\right)-
\dfrac{u_{in_0}(\hat{s}+\alpha\hat{t})+u_{in_0}(2-x-\alpha t)}{2}+
\dfrac{1}{2\alpha}\displaystyle\int\limits_{\hat{s}+\alpha\hat{t}}^{2-x-\alpha t}u_{in_1}(\eta)d\eta.
\end{equation}
Therefore, $u$ is given in $D_T^{(II)}$ by the formula
\begin{equation}\label{Form:uDII}
u(x,t)=w(x,t)+\dfrac{u_0(x+\alpha t)-u_0(\hat{s}+\alpha \hat{t})}{2}+
\dfrac{1}{2\alpha}\displaystyle\int\limits_{\hat{s}+\alpha\hat{t}}^{x+\alpha t}u_{in_1}(\eta)d\eta,
\end{equation}
where $w$, $u_0$, $u_1$ are the functions defined by (\ref{w}), (\ref{uj}).

\begin{remark}
Proceeding as in Remark \ref{remcont} it is easy to show that also the function $u$ defined by (\ref{Form:uDII}) is continuous across the characteristic $x+\alpha t=1$.  
\end{remark}

\begin{remark}\label{Re:Odd}
We notice that in order to obtain the representation formulas (\ref{Form:uDI}), (\ref{Form:uDII}) it was only necessary to extend data $u_{in_0}, u_{in_1}$ as odd functions with respect to $x=1$ (see (\ref{uj})).
\end{remark}

\subsection{Existence and uniqueness of local solution to problem ($\widetilde{\text{P}}_c$)}
Assume for a moment that $s\in C^2[0,T]$, $-\alpha<\dot{s}<0$ and let $u$ be the function given by
\begin{equation}\label{Form:u}
u(x,t)= \left\{ \begin{array}{lcc}
w(x,t)+\dfrac{u_0(x-\alpha t)+u_0(x+\alpha t)}{2}+\dfrac{1}{2\alpha}\bigintsss\limits_{x-\alpha t}^{x+\alpha t}  u_1(\eta)d\eta
 &\quad   \text{in}  & D_T^{(I)} \\
 \\
             w(x,t)+\dfrac{u_0(x+\alpha t)-u_0(\hat{s}+\alpha \hat{t}) }{2}+\dfrac{1}{2\alpha}\bigintsss\limits_{\hat{s}+\alpha \hat{t}}^{x+\alpha t}  u_1(\eta)d\eta &\quad  \text{in} & D_T^{(II)},
             \end{array}
   \right. 
\end{equation}
where $w$ is defined by (\ref{w}), $u_0$, $u_1$ are given by (\ref{uj}) and $\hat{t}$ is the unique solution to equation (\ref{hat-t}). A necessary condition on $u$ to be $(u,s)$ a solution to ($\widetilde{\text{P}}_u$) on $\overline{D_T}$ is $u\in C^{2,2}\left(\overline{D_T}\right)$. To obtain this regularity in $D_T^{(I)}$ and $D_T^{(II)}$ we require
\begin{enumerate}
\item[(H1)]$ \ \ \ \ \ \ \ \ \ \ \ u_{in_0}\in C^2[s_0,1],
 \ \ \ \ \ \ \ \ \ \ \ u_{in_1}\in C^1[s_0,1]$.
\end{enumerate}
We will also require on data the compatibility conditions
\begin{enumerate}
\item[(H2)] $\ \ \ \ \ \ \ \ \ \ \ u_{in_0}(s_0)=0,\ \ \ \ \ \ \ \ \ \ \ u''_{in_0}(1)=0$,
\end{enumerate}  
which assure the function $u$ is continuous on the corner $( s_0,0)$ (first condition) and the equation $u_{tt}=\alpha^2 u_{xx}$ is satisfied in the corner $(1,0)$ (second condition).

We will look now for conditions that ensure the continuity of $u$ and of its partial derivatives on the characteristic curve $\Sigma$ given by 
\begin{equation*}
\Sigma: x-\alpha t=s_0, \quad 0< t< T.
\end{equation*}
The continuity of $u$ across $\Sigma$ can be easily proved from (\ref{Form:u}) by checking that
\begin{equation*}
\displaystyle\lim_{x-\alpha t\to s_0^+}u(x,t)=
\displaystyle\lim_{x-\alpha t\to s_0^-}u(x,t).
\end{equation*}
Taking into account that (\ref{hat-t}) implies
\begin{equation*}
\dfrac{\partial \hat{t}}{\partial t}=-\alpha\dfrac{\partial \hat{t}}{\partial x}=-\dfrac{\alpha}{\dot{s}(\hat{t})-\alpha},
\end{equation*}
that (\ref{w}), (\ref{uj}) yield 
\begin{align*}
&\alpha w_x=w_t=\left\{
\begin{array}{lll}
\mu^{'}\left(\dfrac{x+\alpha t-1}{\alpha}\right)=u_{in_1}(1)&\text{if }&x+\alpha t>1\\
0&\text{if }&x+\alpha t\leq 1,
\end{array}
\right.\\
\\
&u_j^{'}(x)=\left\{
\begin{array}{lll}
u^{'}_{in_j}(x)&\text{if }&s_0\leq x\leq 1\\
u_{in_j}^{'}(2-x)&\text{if }&1<x\leq 2-s_0
\end{array}
\right.\hspace{2cm}j=0,1,
\end{align*}
and $w_{xx}=w_{tt}=0$ since $\mu$ is a linear function of its argument, the following partial derivatives of $u$ can be obtained from (\ref{Form:u}):

\begin{equation}
\label{Form:u_t}
u_t= \left\{ \begin{array}{lcl}
             w_t(x,t)-\dfrac{\alpha}{2}\left(u_0'(x-\alpha t)-u_0'(x+\alpha t) \right)+\\[0.4cm]
             \quad\dfrac{1}{2}\left( u_1(x+\alpha t)+u_1(x-\alpha t)\right)  \ \ \ \ &\textrm{in}& D_T^{(I)} \\ 
\\
             w_t(x,t)+\dfrac{\alpha}{2}\left(u_0'(x+\alpha t)+u_0'(\hat{s}+\alpha \hat{t})\left( \dfrac{\widehat{\dot{s}}+\alpha }{\widehat{\dot{s}}-\alpha }\right) \right)+ \\[0.4cm]
             \quad+\dfrac{1}{2}\left(u_1(x+\alpha t)+u_1(\hat{s}+\alpha \hat{t})\left( \dfrac{\widehat{\dot{s}}+\alpha }{\widehat{\dot{s}}-\alpha }\right) \right) \ \ \ \ \ \ &\textrm{in} &D_T^{(II)}
             \end{array}
   \right.
\end{equation}
\begin{equation}
\label{Form:u_x}
u_x= \left\{ \begin{array}{lcl}     
             w_x(x,t)+\dfrac{1}{2}\left(u_0'(x-\alpha t)+u_0'(x+\alpha t) \right)+\\[0.4cm]
             \quad\dfrac{1}{2\alpha}\left( u_1(x+\alpha t)-u_1(x-\alpha t)\right) \ \ \ \ \ &\textrm{in}& D_T^{(I)} \\
\\
             w_x(x,t)+\dfrac{1}{2}\left(u_0'(x+\alpha t)-u_0'(\hat{s}+\alpha \hat{t})\left( \dfrac{\widehat{\dot{s}}+\alpha }{\widehat{\dot{s}}-\alpha }\right) \right)+ \\[0.4cm]
             \quad+\dfrac{1}{2\alpha}\left(u_1(x+\alpha t)-u_1(\hat{s}+\alpha \hat{t})\left( \dfrac{\widehat{\dot{s}}+\alpha }{\widehat{\dot{s}}-\alpha }\right) \right)\ \ \ \ \ &\textrm{in}&  D_T^{(II)}
             \end{array}
   \right.
\end{equation}

\begin{equation}
\label{Form:u_tt}
u_{tt}= \left\{ \begin{array}{lcl}         
             \dfrac{\alpha^2}{2}\left(u_0''(x-\alpha t)+u_0''(x+\alpha t) \right)+\\[0.4cm]
             \quad\dfrac{\alpha}{2}\left( u_1'(x+\alpha t)-u_1'(x-\alpha t)\right)             
              \ \ \ \ \ &\textrm{in}&  D_T^{(I)}\\[1cm]               
             \dfrac{\alpha^2}{2}\left(u_0''(x+\alpha t)-u_0''(\hat{s}+\alpha \hat{t})\left( \dfrac{\widehat{\dot{s}}+\alpha }{\widehat{\dot{s}}-\alpha }\right)^2+\right.\\[0.6cm]
               \quad\left.\dfrac{2\alpha\widehat{\ddot{s}}u_0'(\hat{s} +\alpha \hat{t})}{\widehat{\dot{s}}-\alpha)^3} \right)+ \\[0.4cm]
			\quad+\dfrac{\alpha}{2}\left(u_1'(x+\alpha t)-u_1'(\hat{s}+\alpha \hat{t})\left( \dfrac{\widehat{\dot{s}}+\alpha }{\widehat{\dot{s}}-\alpha }\right)^2+\right.\\[0.6cm]
			\quad\left. \dfrac{2\alpha\widehat{\ddot{s}}u_1(\hat{s}+\alpha\hat{t})}{(\widehat{\dot{s}}-\alpha)^3} \right)\ \ \ \ \ &\textrm{in}&  D_T^{(II)}
             \end{array}
   \right.
\end{equation}

\begin{equation}
\label{Form:u_xx}
u_{xx}= \left\{ \begin{array}{lcl}         
             \dfrac{1}{2}\left(u_0''(x-\alpha t)+u_0''(x+\alpha t) \right)+\\[0.4cm]
             \quad\dfrac{1}{2\alpha}\left( u_1'(x+\alpha t)-u_1'(x-\alpha t)\right)             
              \ \ \ \ \ &\textrm{in} D_T^{(I)} \\[1cm]               
             \dfrac{1}{2}\left(u_0''(x+\alpha t)-u_0''(\hat{s}+\alpha \hat{t})\left( \dfrac{\widehat{\dot{s}}+\alpha }{\widehat{\dot{s}}-\alpha }\right)^2+\right.\\[0.6cm]
               \quad\left.\dfrac{2\alpha\widehat{\ddot{s}}u_0'(\hat{s}) +\alpha \hat{t})}{(\widehat{\dot{s}}-\alpha)^3} \right)+ \\[0.4cm]
			\quad+\dfrac{1}{2\alpha}\left(u_1'(x+\alpha t)-u_1'(\hat{s}+\alpha \hat{t})\left( \dfrac{\widehat{\dot{s}}+\alpha }{\widehat{\dot{s}}-\alpha }\right)^2+\right.\\[0.6cm]
			\quad\left. \dfrac{2\alpha\widehat{\ddot{s}}u_1(\hat{s}+\alpha\hat{t})}{(\widehat{\dot{s}}-\alpha)^3} \right)\ \ \ \ &\textrm{in}&D_T^{(II)}
             \end{array}
   \right.
\end{equation}
where we have set $\widehat{\dot{s}}=\dot{s}(\hat{t})$ and $\widehat{\ddot{s}}=\ddot{s}(\hat{t})$.

\begin{remark} We observe that the first and second derivatives of $u(x,t)$ are continuous across the characteristic 
$x+\alpha t=1$.
\end{remark}
\noindent From (\ref{Form:u_t})-(\ref{Form:u_x}) we find that $u_t$ and $u_x$ are continuous across $\Sigma$ if and only if
\begin{enumerate}
\item[(A)] $\dot{s}_0 u_{in_0}'(s_0)+u_{in_1}(s_0)=0$.
\end{enumerate}
Similarly, from (\ref{Form:u_tt}),  (\ref{Form:u_xx}) we find that the continuity of $u_{tt}$ and $u_{xx}$ across $\Sigma$ is accomplished if and only if
\begin{enumerate}
\item[(B)] $(\dot{s}_0-\alpha)\left[(\dot{s}_0^2+\alpha^2)u''_{in_0}(s_0)+2\dot{s}_0 u'_{in_1}(s_0) \right]-\ddot{s}_0\left[\alpha u'_{in_0}(s_0)+u_{in_1}(s_0) \right]=0$,
\end{enumerate}
where $\ddot{s}_0=\ddot{s}(0)$. 

Thus, if (H1), (H2), (A), (B) hold, $s\in C^2[0,T]$ with $-\alpha<\dot{s}<0$, and $u$ is defined by (\ref{Form:u}) then $(u,s)$ satisfies the first five conditions of ($\widetilde{\text{P}}_u$) on the time interval $[0,T]$. In the following we will focus on the last two conditions of problem ($\widetilde{\text{P}}_u$). 

Evaluating (\ref{Form:u_x}) in $(s,t)$ and noting that $\hat{t}(s,t)=t$, we find
\begin{equation}\label{Form:u_x-s}
u_x(s,t)=\frac{\alpha \left(u_0\right)'(s+\alpha t)+u_1(s+\alpha t)}{\alpha-\dot{s}}.
\end{equation}
Since $-\alpha<\dot{s}<0$ implies $s_0\leq s+\alpha t\leq 1$, the functions $u_0'$, $u_1$ in (\ref{Form:u_x-s}) are equal to  $u'_{in_0}$, $u_{in_1}$. The free boundary $s$ must hence solve the Cauchy problem
\begin{equation}\label{Eq:s-1}
(\text{P}_s)\,:\,
\left\{\begin{array}{l}
-\dot{s}=
\left(\dfrac{g(s+\alpha t)}{\alpha-\dot{s}}-\beta\right)_+\hspace{1cm}0<t<T\\
\\
s(0)=s_0,\end{array}\right.
\end{equation}
where we have set
\begin{equation*}
\beta=\frac{\lambda+\delta}{2\lambda},
\hspace{2cm}
2\lambda g(x)=\alpha u'_{in_0}(x)+u_{in_1}(x)\hspace{0.75cm} s_0\leq x\leq 1.
\end{equation*}
Then, condition $-\alpha<\dot{s}<0$ is equivalent to 
\begin{equation*}
0<\dfrac{g(s+\alpha t)}{\alpha-\dot{s}}-\beta<\alpha \ \ \ \ \ \ \ \ \ \ \ \ \text{i.e.} \ \ \ \ \ \ \ \ \ \ \
\beta<\dfrac{g(s+\alpha t)}{\alpha-\dot{s}}<\alpha+\beta. 
\end{equation*}
Therefore we require
\begin{equation*}
\beta<\inf_{[0,T]} \left\{\dfrac{g(s+\alpha t)}{\alpha-\dot{s}}\right\}\leq \sup_{[0,T]} \left\{\dfrac{g(s+\alpha t)}{\alpha-\dot{s}}\right\}<\alpha+\beta,
\end{equation*}
yielding 
\begin{equation}
\beta< \dfrac{\inf_{[s_0,1]} g(x)}{2\alpha}\leq \dfrac{\sup_{[s_0,1]}g(x)}{\alpha}<\alpha+\beta. 
\end{equation}
Then we get the following condition
\begin{enumerate}
\item[(H3)] $\ \ \ \ \ \ \ \ \ \ \ \ 2\alpha\beta<\inf_{[s_0,1]} g(x), \ \ \ \ \ \ \ \ \ \ \sup_{[s_0,1]} g(x)<\alpha \beta+\alpha^2$,
\end{enumerate}
which is consistent only if 
\begin{enumerate}
\item[(H4)] $\ \ \ \ \ \ \ \alpha>\beta$.
\end{enumerate}
In conclusion, if (H3), (H4) are satisfied then $-\alpha<\dot{s}<0$ and  equation (\ref{Eq:s-1})$_1$ 
can be rewritten as 
\begin{equation}\label{Eq:s-2}
\dot{s}^2-(\alpha+\beta)\dot{s}-\left[g(s+\alpha t)-\alpha \beta\right]=0.
\end{equation}
It is easy to see that equation (\ref{Eq:s-2}) is an algebraic second order equation in $\dot{s}$ with 
two distinct real roots, with only one of them negative given by
\begin{equation}\label{Eq:s-3}
\dot{s}=\left(\dfrac{\alpha+\beta}{2}\right)-\sqrt{\left(\dfrac{\alpha+\beta}{2}\right)^2+
\Big[g(s+\alpha t)-\alpha \beta\Big]}.
\end{equation}  

On one hand, recalling (A) and (\ref{Eq:s-1}) \footnote{To avoid a too heavy notation here  $u_{in_0}$,  $u_{in_1}$ 
and their derivatives represent the value of these functions evaluated in $s_0$.},
\begin{equation}\label{Eq:s-1bis}
\left\{\begin{array}{l}
-\dot{s}_0=
\left(\dfrac{1}{2\lambda}\dfrac{\alpha u'_{in_0}+u_{in_1}}{\alpha-\dot{s}_0}-\beta\right)\\
\\
\dot{s}_0 =-\dfrac{u_{in_1}}{u_{in_0}'}.
\end{array}\right.
\end{equation}
On eliminating $\dot{s}_0$ from (\ref{Eq:s-1bis}) we find 
\begin{enumerate}
\item[(H5)]
$\ \ \ \ \ \ \ \ \ \ u_{in_1}=u'_{in_0}\left(\dfrac{u'_{in_0}}{2\lambda}-\beta\right)$,
\end{enumerate}
which provides the condition that guarantees the continuity of the first derivatives of $u$ across $\Sigma$. 

On the other hand, from (A) we see that $u_{in_1}=-\dot{s}_0 u_{in_0}'$. Replacing this in (B) we find
\begin{equation}
(\dot{s}_0-\alpha)\left[(\dot{s}_0^2+\alpha^2)u''_{in_0}+2\dot{s}_0 u'_{in_1} \right]=
\ddot{s}_0u'_{in_0}(\alpha-\dot{s}_0),
\end{equation}
which simplifies to
\begin{equation*}
-\left[(\dot{s}_0^2+\alpha^2)u''_{in_0}+2\dot{s}_0 u'_{in_1} \right]=
\ddot{s}_0u'_{in_0}.
\end{equation*}
Replacing (\ref{Eq:s-1bis})$_2$ in the above, after some algebra, we find
\begin{equation}\label{esse2}
\ddot{s}_0=-\dfrac{1}{(u'_{in_0})^3}\left[u{''}_{in_0}(u^2_{in_1}+
u^{'^{2}}_{in_0}\alpha^2)-2u_{in_1} u'_{in_1}u'_{in_0}\right]
\end{equation}
Now let us go back to equation (\ref{Eq:s-3}). Differentiating with respect to time we get 
\begin{equation*}
\ddot{s}=-\dfrac{g^{'}(s+\alpha t)(\dot{s}+\alpha)}{\sqrt{\left(\alpha+\beta\right)^2+
4\Big[g(s+\alpha t)-\alpha \beta\Big]}},
\end{equation*}
or equivalently 
\begin{equation*}
\ddot{s}=\dfrac{g^{'}(s+\alpha t)(\dot{s}+\alpha)}{2\dot{s}-(\alpha+\beta)}.
\end{equation*}
Therefore, recalling (\ref{Eq:s-1bis})$_2$ and the definition of $g$, we find  
\begin{equation}\label{qui}
\ddot{s}_0=\dfrac{g^{'}(s_0)(\dot{s}_0+\alpha)}{2\dot{s}_0-(\alpha+\beta)}=
\dfrac{(\alpha u^{''}_{in_0}+u'_{in_1})(u_{in_1}-\alpha u'_{in_0})}{2\lambda(2u_{in_1}+(\alpha+\beta)u'_{in_0})}.
\end{equation}
Eliminating $\ddot{s}_0$ between (\ref{esse2}) and (\ref{qui})
we find 
\begin{enumerate}
\item[(H6)]
$\dfrac{2u_{in_1} u'_{in_1}u'_{in_0}-u{''}_{in_0}(u^2_{in_1}+
u^{'^{2}}_{in_0}\alpha^2)}{(u'_{in_0})^3}=
\dfrac{(\alpha u^{''}_{in_0}+u'_{in_1})(u_{in_1}-\alpha u'_{in_0})}{2\lambda(2u_{in_1}+(\alpha+\beta)u'_{in_0})}$,
\end{enumerate}
which provides the condition that guarantees the continuity of the second derivatives of $u$ across $\Sigma$.

We finally observe that if $s$ satisfies (\ref{Eq:s-3}) then conditions (H1), (H3) imply $s\in C^2[0,T]$. 
Therefore, we can establish the following characterization result for any solution to problem ($\widetilde{\text{P}}_u$):

\begin{lemma}\label{Le:Charact-1}
If $T$ satisfies (\ref{T}) and (H1)-(H6) hold, $s$ is a solution to the initial value problem (\ref{Eq:s-1}), and $u$ is defined by (\ref{Form:u}) then $(u,s)$ is a classical solution to problem ($\widetilde{\text{P}}_u$) in the time interval $[0,T]$. Moreover, $s\in C^2[0,T]$, $-\alpha<\dot{s}<0$ and $\dot{s}(0)=\dot{s}_0$ with $\dot{s}_0$ defined by (\ref{Cond:s1_Compatibility-1}).
\end{lemma}

Classical results on ordinary differential equations together with conditions (H1), (H3) assure that problem (P$_s$) has a unique solution $s$ in the interval $[0,T]$. Then Lemma \ref{Le:Charact-1} assures the existence of local solutions to ($\widetilde{\text{P}}_u$). Finally, since the characterization of $u$ and $s$ established by Lemma \ref{Le:Charact-1} only involves data in problem ($\widetilde{\text{P}}_u$), uniqueness of solution to (P$_s$) implies uniqueness of the local solution to ($\widetilde{\text{P}}_u$) in the time interval $[0,T]$. Thus, we have the following

\begin{theorem}
If $T$ satisfies (\ref{T}) and (H1)-(H6) hold, then problem ($\widetilde{\text{P}}_c$) has a unique local solution $(c,s)$ in the time interval $[0,T]$. Moreover, $c$ is given by (\ref{c-u_Transformation}) with $u$ defined by (\ref{Form:u}), and $s$ is the unique solution to the Cauchy problem ($\text{P}_s$). 
\end{theorem}

\subsection{Continuation of the solution: existence and uniqueness in the large}

Let us set

\begin{equation*}
T_0=\dfrac{1-s_0^{(0)}}{\alpha}, \ \ \ \ \ \ \ \ \  s_0^{(0)}=s_0, \ \ \ \ \ \ \ \ \ \ \ \ \ s^{(0)}(t)=s(t),
\end{equation*}
where $s$ is the free boundary in the time interval $[0,T_0]$. Since $s_0>0$, there are two possible situations: 

\begin{itemize}
\item [(a1)]  $s_0^{(0)}(t)=0$ for some time $t<T_0$,
\item [(a2)]  $s_0^{(0)}(t)>0$ for all times $t\in [0,T_0]$.
\end{itemize}
If (a1) holds the solid part is completely consumed in a finite time $T_{fin}<T_0$. From time $T_{fin}$ on,
our model is no longer a free boundary problem and the evolution of $u$ is described by the wave equation 
in the strip $[0,1]$. If (a2) holds, then the solution can be extended to the  interval $[T_0,T_1]$ with 
\begin{equation*}
T_1=T_0+\dfrac{1-s_0^{(1)}}{\alpha}, \ \ \ \ \ \ \ \ \  s_0^{(1)}=s^{(0)}(T_0)<s_0^{(0)}.
\end{equation*}
The formal expression for $u$ is obtained again through the D'Alembert formula extending the initial data in the same odd fashion for $x>2-s_0$ (see Remark \ref{Re:Odd}). The equation for $s$ thus remains the same for $t>T_0$. We denote the free boundary in the time interval $[T_0,T_1]$ by $s^{(1)}(t)$. Again we distinguish between the case in which $s^{(1)}(t)=0$ for some $t\in (T_0,T_1]$ and the case in which $s^{(1)}(t)>0$ for all $t\in (T_0,T_1]$. Proceeding in this way we can build the sequence
\begin{equation*}
0<T_{j+1}=T_j+\dfrac{1-s_0^{(j+1)}}{\alpha}, \ \ \ \ \ \ \ \ \ \ s_0^{(j+1)}<s_0^{(j)},
\end{equation*}
which of course makes sense if $s^{(j)}(t)>0$ for all $j$ (otherwise the solid part is totally consumed in a finite time $T_{fin}<T_0$). We have that
\begin{equation*}
\displaystyle\lim_{j\to+\infty}T_j=+\infty.
\end{equation*}
Indeed,
\begin{equation*}
\displaystyle\lim_{j\to+\infty}T_j=
\displaystyle\lim_{j\to+\infty}\left(T_0+\displaystyle\sum_{k=0}^j\left(
T_{k+1}-T_k\right)\right)
=T_0+\displaystyle\sum_{k=0}^{\infty}\dfrac{1-s_0^{(k)}}{\alpha},
\end{equation*}
with
\begin{equation*}
\displaystyle\sum_{k=0}^{\infty}\dfrac{1-s_0^{(k)}}{\alpha}=+\infty
\hspace{1cm}\text{since}\hspace{1cm}
\displaystyle\lim_{j\to\infty}\dfrac{1-s_0^{(k)}}{\alpha}\geq\dfrac{1-s_0}{\alpha}>0.
\end{equation*}

Therefore our model predicts one of the following situations:
\begin{enumerate}
\item[(i)] the marble is never completely consumed, being the evolution of its surface described by $s$ and the concentration of $H^+$ ions described by $c$, with $(c,s)$ the unique global solution to problem ($\widetilde{\text{P}}_c$),
\item[(ii)] the marble is completely consumed in a finite time $T_{fin}>0$, being the system described by:
\begin{enumerate}
\item[(a)] the unique solution $(c,s)$ to problem ($\widetilde{\text{P}}_c$) in the time interval $[0,T_{fin}]$,
\item[(b)] the solution $c$ of an initial-boundary value problem for the wave equation ($\widetilde{\text{P}}_c$)$_1$ in the strip $[0,1]$ and the time interval $(T_{fin},+\infty)$ ($s\equiv0$ from $T_{fin}$ on).
\end{enumerate}
\end{enumerate}
We notice that in the second case (ii), only the first option (a) is physically relevant.

\section{Simulations}

In this section we simulate the behavior of the free boundary by solving numerically problem $(\text{P}_s)$. We begin by a simple case in which the solution can be determined analytically.

\subsection{The simple case $g=const$}

Suppose 
\begin{equation*}
u_{in_0}=A(x-s_0), \ \ \ \ \ \ \ \ \ \ \ \ \ \ u_{in_1}=A\left(\dfrac{A}{2\lambda}-\beta\right),
\end{equation*}
with $A$ a positive constant. In this case $g$ is constant and hence the free boundary $x=s(t)$
is a line. It is easy to show that conditions (H1), (H2) are automatically fulfilled. Taking the values 
$$A=4, \ \ \ \ \ \lambda=1, \ \ \ \ \ \ \ \ \alpha=3, \ \ \ \ \ \ \ \ \ \delta=10^{-5},\ \ \ \ \ \ \ \ \ s_0=0.25,$$
we find that also conditions (H3)-(H6) hold and the free boundary is 
\begin{equation*}
s(t)=s_0+\dot{s}_0 t,
\end{equation*} 
with
\begin{equation*}
\dot{s}_0=-\dfrac{1}{2\lambda}\left(A-\lambda-\delta\right)_+.
\end{equation*}

\begin{figure}[h!]
\centering
\includegraphics[scale=0.5]{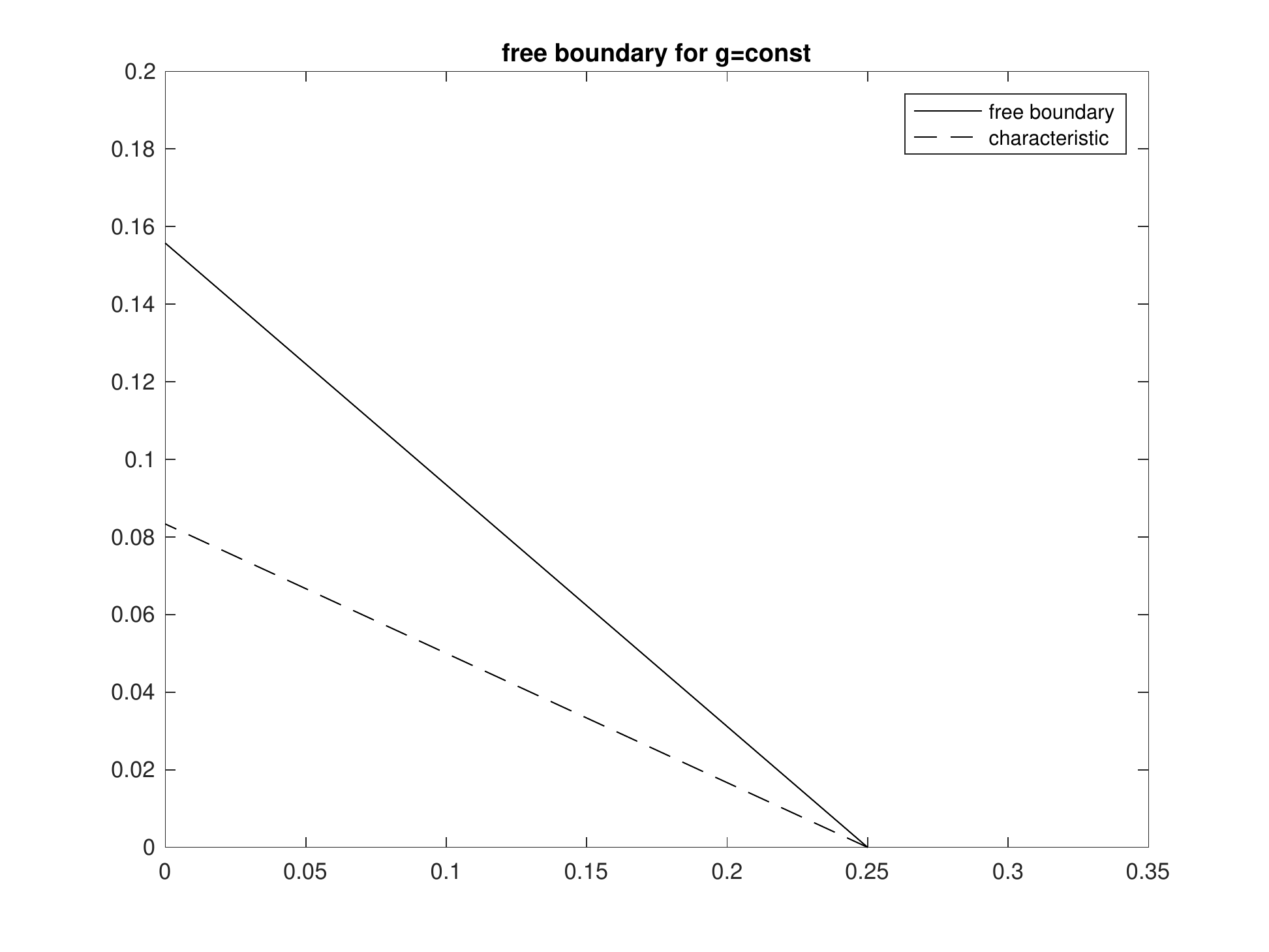}
\caption{Free boundary $x=s(t)$ for $g=const$.}
\label{Fig:line}
\end{figure}

The plot of the free boundary $x=s(t)$ and of the characteristic curve $x+\alpha t=s_0$ are shown in Fig. \ref{Fig:line}. In this case 
the solid part of the domain vanishes in a finite time given by 
\begin{equation*}
T_{fin}=-\dfrac{s_0}{\dot{s}_0}\approx 0.166< \dfrac{(1-s_0)}{\alpha}=0.25.
\end{equation*}

\subsection{The case of a non constant  $g$}

Assume that

\begin{equation*}
u_{in_0}=A(x-s_0)(1-x)^3,\ \ \ \ \ \ \ \ \ \ \ \ \ \
u_{in_1}=B(1-x)^{3/2},
\end{equation*}
so that
\begin{equation*}
u_{in_0}^{'}=-A(1-x)^2(4x-3s_0-1), \ \ \ \ \ \ \ \ \ u_{in_0}^{''}=6A(1-x)(2x-s_0-1).
\end{equation*}
Notice that hypothesis (H1), (H2) are satisfied. In this case  (H4), (H5) are not fulfilled so that the function $u$ is continuous up to the second derivatives everywhere except on the characteristic $\Sigma$ where the derivatives of $u$ experience a jump discontinuity. 

\begin{figure}[h!]
\centering
\includegraphics[scale=0.5]{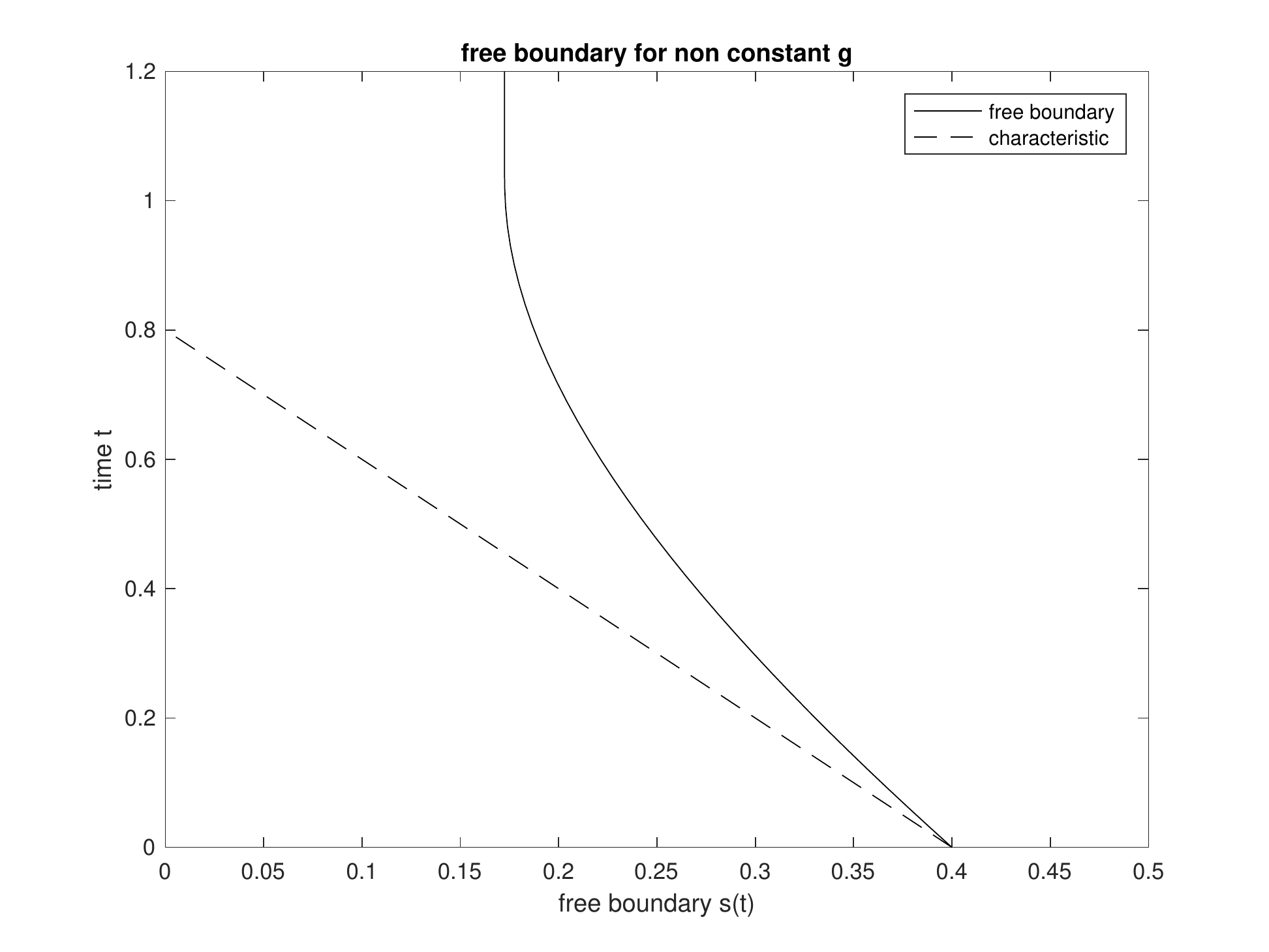}
\caption{Free boundary $x=s(t)$ for non constant $g$.}
\label{Fig:nonlin}
\end{figure}

Taking the values 
$$A=1, \ \ \ \ \  B=3, \ \ \ \ \ \  \lambda=1, \ \ \ \ \ \ \ \ \alpha=0.5, \ \ \ \ \ \ \ \ \ \delta=10^{-5}, \ \ \ \ \ \ \ \ \ s_0=0.4,$$
and solving the problem ($\text{P}_s$) we get the solution of Fig. \ref{Fig:nonlin}. We notice that this solution satisfies $-\alpha<\dot{s}<0$.

In this case we observe that $\dot{s}=0$ for $t\geqslant 1.08$, which means that the reaction ceases at time $t=1.08$ and that the initial data are such that the solid slab is not completely worn away by the acid solution. 

\begin{figure}[h!]
\centering
\includegraphics[scale=0.5]{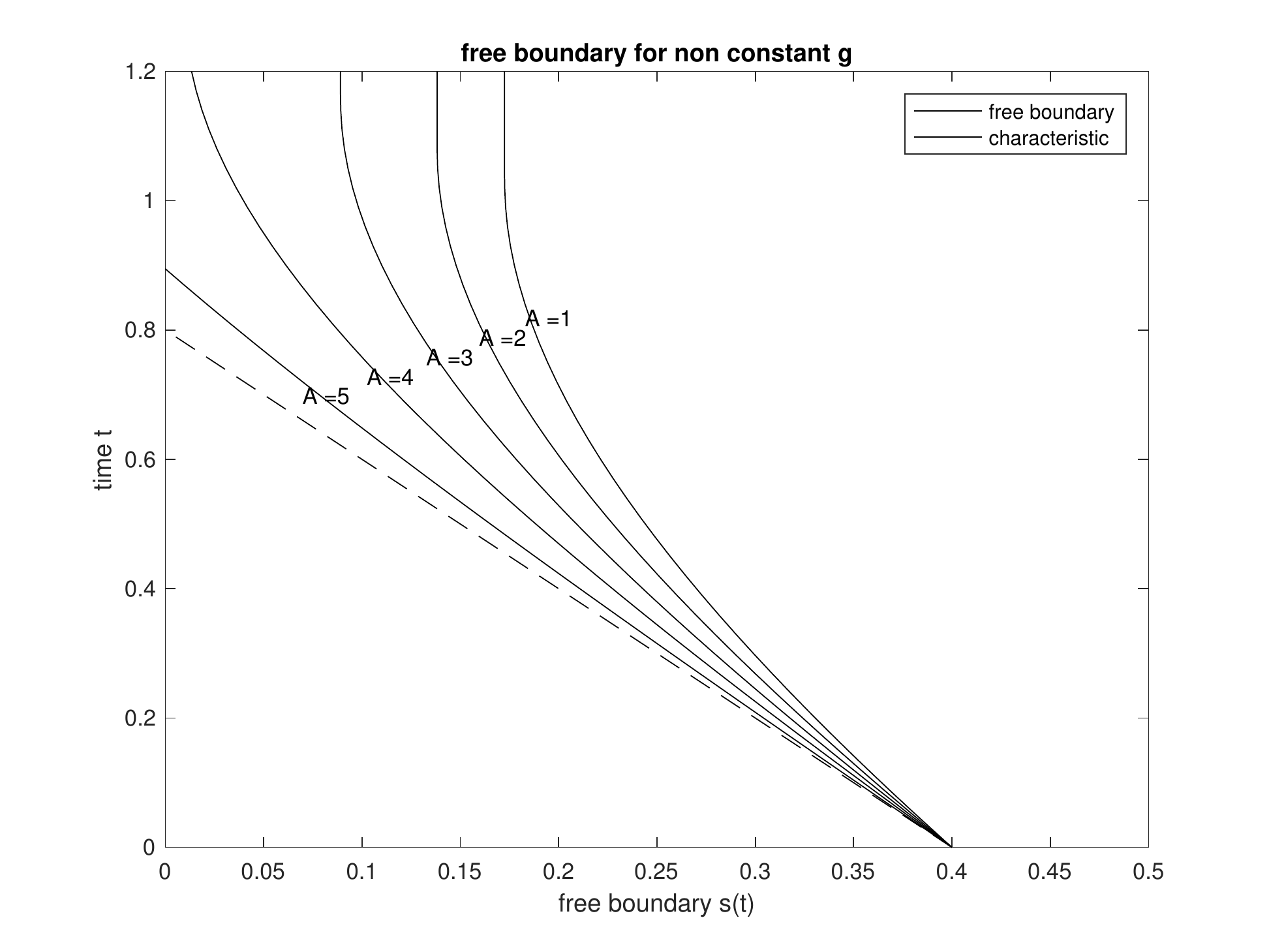}
\caption{Free boundary $x=s(t)$ for non constant $g$ and $A=1,2,3,4,5$.}
\label{Fig:nonlin2}
\end{figure}

If we increase the value of $A$ (which corresponds to increase the initial concentration of H$^{+}$ ions), the ``asymptotic
value'' of $s$ - i.e. the thickness of the slab when $\dot{s}$ vanishes - becomes smaller, as shown in Fig. \ref{Fig:nonlin2}.
This is physically consistent since a more acid solution is expected to erode a larger part of the slab. 

\noindent

\section{Conclusion}
In this paper we present a mathematical model for the diffusive-reactive process between calcium carbonate ($\mathrm{CaCO}_3$) and an aqueous solution containing sulphuric acid ($\mathrm{H}_2\mathrm{SO}_4$). We model the case of a marble slab immersed in an acid solution with the reaction taking place on the contact surface between the slab and the liquid. The system is described by the evolution of both the H$^{+}$ ions concentration in the liquid and the free boundary $s$ where the reaction occurs. We assume the H$^{+}$ ions diffuse according to Cattaneo's law and formulate the mathematical problem in a one dimensional geometry writing the molar mass balance, the reaction kinetics and the net flux of H$^{+}$ ions on the reacting surface. The problem turns out to be a hyperbolic moving boundary problem for the telegraph equation in which the free boundary represents the thickness of the slab that is consumed because of the reaction. Using typical values taken from the literature, we show that the diffusive time scale is smaller than the reaction time scale, so that the problem can be reduced to a free boundary problem for the wave equation. We also write representation formulas for the $\mathrm{H}^+$ ions concentration and show that the evolution of the free boundary is given by a nonlinear differential equation involving the initial data of the problem. Under appropriate hypotheses on the data we prove existence and uniqueness in the large and we determine conditions that guarantee the regularity of the solution. Finally we perform some numerical simulations to illustrate the behavior of the free boundary.  

\section*{Acknowledgement}
This paper has been partially sponsored by Universidad Nacional de Rosario and CONICET (Argentina).

\bibliographystyle{plain} 
\bibliography{References}

\end{document}